# Multiple-Particle Autofocusing Algorithm Using Axial Resolution and Morphological Analyses Based on Digital Holography

Wei-Na Li,* Yi Zhou, Jiatai Chen, Hongjie Ou and XiangSheng Xie*

Physics Department, College of Science, Shantou University, Shantou, Guangdong, 515065, China
22yzhou7@stu.edu.cn (Y.Z.); 22jtchen4@stu.edu.cn (J.C.); 22hjou@stu.edu.cn (H.O.);
*Correspondence:weinali@stu.edu.cn; xxs@stu.edu.cn;

**Abstract:** We propose an autofocusing algorithm to obtain, relatively accurately, the 3D position of each particle, particularly its axial location, and particle number of a dense transparent particle solution via its hologram. First, morphological analyses and constrained intensity are used on raw reconstructed images to obtain information on candidate focused particles. Second, axial resolution is used to obtain the real focused particles. Based on the mean intensity and equivalent diameter of each candidate focused particle, all focused particles are eventually secured. Our proposed method can rapidly provide relatively accurate ground-truth axial positions to solve the autofocusing problem that occurs with dense particles.

## 1. Introduction

Digital holography (DH) is an advanced optical technique designed to record and reconstruct optical information on three-dimensional (3D) objects. Unlike conventional microscopy, which focuses on only a single plane, digital holographic microscopy (DHM) can capture a volume, and reconstruct every plane of this volume. Currently, sizing, counting, and locating problems related to in situ micro-objects (bubbles, particles, or microorganisms) have gained significant attention from researchers, particularly with the development of in-line DH and its potential as an alternative to conventional microscopy [1,2]. In DH and DHM, autofocusing is used to obtain the exact location of an object. Although autofocusing can be achieved using experimental configurations [3,4], it is realized by means of computation algorithms, such as image sharpness [5,6], structure tensor [7], edge sparsity [8], and magnitude differential [9].

A number of researchers [10-13] have employed spherical waves to illuminate particles suspended in water. They determined the position and measured the size of each particle using a captured hologram. The experimental configuration provided a magnification of the target object. However, this approach reduces the field of view (FOV); furthermore, the accuracy of the number of particles and their locations along the z-axis are significantly affected by the chosen depth spacing. On the other hand, Tian *et al.* investigated bubbles in water [14] and utilized the minimum intensity as a focus metric to detect the edges of the bubbles, thereby determining the location of each bubble, particularly its axial position. However, although the processing speed of their proposed approach is faster, the location information obtained is inaccurate. Moreover, when several bubbles are clustered together, they are erroneously recognized as a single bubble. These conventional diffraction-based methods cause severe defocused-image problems if the ground-truth z-position of each micro-object is unknown, which results in inaccuracies regarding the micro-objects on the mount and their locations. Later on, Lang *et al.* [15] utilized the Q value as the focus metric to recognize the best axial location for plankton; however, this focus metric is not suitable for numerous particles, especially dense particles in DH.

Compressive models with sparsity have exhibited good performance on noise and ghost images by transforming the hologram reconstruction problem into a regularized nonlinear optimization. Brady *et al.* introduced a compressive sensing algorithm for DH, and demonstrated that decompressive inference can infer multidimensional objects from a 2D hologram [16]. Liu *et al.* applied compressive holography to object localization [17,18], which significantly improves the accuracy of lateral localization as long as the solution is sparse in its derivatives. Chen *et al.* used a plane wave to illuminate bubbles [19], and proposed a compressive holographic method to locate the axial position of each bubble; however, their proposed method is unable to distinguish bubbles that are completely or partially overlapped along the z-axis, nor could it process dense particles. In all of the aforementioned studies, a common approach followed by the researcher was to use total variation regularizers. By contrast, Li *et al.* [20] applied a 3-D hybrid-Weickert nonlinear diffusion regularizer to DH, which can determine the locations of certain small-sized transparent scattering particles that overlap on the z-axis. Consequently, similar autofocusing for multiple micro-objects was achieved while simultaneously removing defocused images. However, all of the aforementioned methods have their drawbacks, including slow processing speed, inability to process dense particles due to a sparse prior, and difficulty in fine-tuning the parameters, among others.

With the recent development of machine learning, Ren *et al.* [21] efficiently employed convolutional neural networks (CNN) to designate autofocusing as a classification problem and provide approximations of the focusing distance for each classification. This approach is more appropriate for single large objects, although reconstructed images are not required. Lee *et al.*[22] proposed first determining the centroid of each particle and then feeding the cropped hologram of each particle into a CNN to obtain the depth information of each particle. Shao *et al.* [23] and Wu *et al.* [24] used a modified U-net network to obtain 3D morphology information of all particles (including the 3D position, size, and shape of each particle), mainly from holograms. Li *et al.* [25] and Hao *et al.* [26] combined Dense Block with U-net to obtain a 3D particle distribution with particle sizes particularly from reconstructed images generated from holograms. Li *et al.* [27] and Ou *et al.* [28] utilized a modified CNN and ResNet to predict the particle number from holograms and raw reconstructed images, which had been inaccurate thus far, to obtain the 3D position of each particle, especially the axial position.

In this study, we propose an autofocusing method based on morphological analyses and axial resolution to obtain, relatively accurately, the 3D position of each particle, particularly its axial location, and the particle number of a dense transparent particle solution via a hologram. Our proposed method has two main components. First, morphological analyses and constrained intensity are used on raw reconstructed images, from which information on candidate focused particles is obtained and saved in one matrix. Second, axial resolution is used to obtain the real focused particles from the aforementioned matrix. For the focus metric, we propose using the product of the mean intensity and equivalent diameter of each candidate focused particle. Eventually, we are able to secure all focused particles for one hologram. In our experiments, we used particles located both at fixed distances and in a particle solution filling a cuvette to examine and verify the proposed method.

The remainder of this paper is organized as follows. Section 2 introduces the principles. Section 3 presents the methodology, in which the description of the algorithm is a key point. Section 4 discusses the experimental results and analyses. Finally, Section 5 presents the conclusions of our study.

## 2. Principles

Herein, it is assumed there are a large number of transparent particles suspended in Milli-Q water, all with a uniform size (diameter) denoted by $pcle_i(\xi,\eta)$, where i = 1, 2, ... , m. A plane wave with wavelength $\lambda = 532nm$ illuminates these particles, and holograms of the 3D information of all the particles in the entire volume are captured using a complementary metal oxide semiconductor (CMOS) camera, as shown in Fig. 1. If each particle is suspended at a distance $z_i$ from the image sensor chip (hologram plane), the Fresnel diffraction [29] for each particle can be mathematically expressed as

$$H_i(x,y,z_i) = FT^{-1}\left\{\exp(jkz_i) \times FT\left\{pcle_i(\xi,\eta)\right\} \times \exp\left(-j\pi\lambda z_i\left(f_x^2 + f_y^2\right)\right)\right\}, \quad (1)$$

$$H(x,y) = \sum_i^m H_i(x,y,z_i), \quad (2)$$

$$I(x,y) = |R(x,y)|^2 + |H(x,y)|^2 + R^*(x,y)H(x,y) + R(x,y)H^*(x,y) + n, \quad (3)$$

$$I_{final}(x,y) = H(x,y) + nse. \quad (4)$$

where $\{\xi,\eta\}$ denotes the lateral coordinates of the particle position, $(f_x, f_y)$ signifies the spatial frequency domain, $(x,y)$ represents the hologram plane, $H(x,y)$ and $H^*(x,y)$ denote the complex amplitude and conjugate, respectively, of the hologram $I(x,y)$, and n indicates the noise induced by the optical system (including high-frequency speckle noise and diffraction patterns of dust or bubbles in the optical path, among others). Owing to the use of a plane wave as the reference beam, the amplitude of the reference beam $R(x,y)=1$. Therefore, the hologram can be rewritten as shown in Eq. (4), in which $nse = 1 + |H(x,y)^2| + H^*(x,y) + n$.

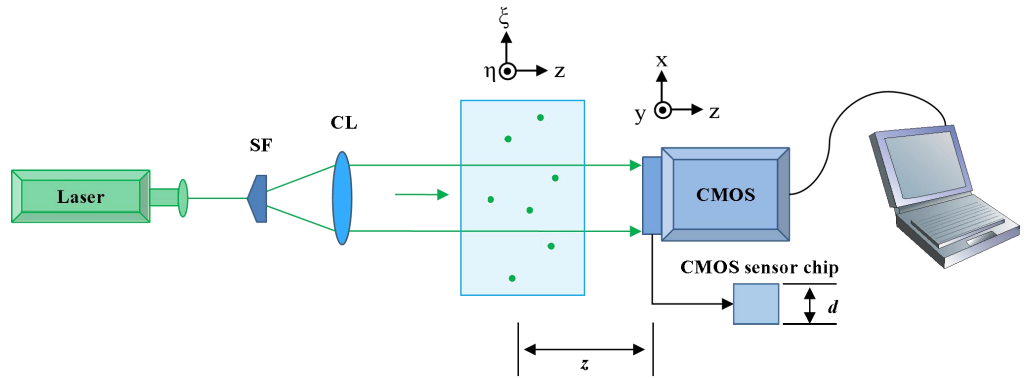

**Figure 1.** Schematic diagram of in-line digital holographic setup for capturing holograms of multiple particles (SF: spatial filter, CL: collimating lens, z: transmission distance between object plane and hologram plane, d: height of CMOS sensor chip).

## 3. Methodology

*3.1. Axial Resolution*

In digital holographic system, assume a point light source with a wavelength $\lambda$ is located at a distance $z$ away from the CMOS sensor (hologram plane) whose size is $hy \times hx$ and pixel pitch is $\Delta_{pp}$, therefore, the height of the hologram is $D = hy \cdot \Delta_{pp}$. The schematic diagram is depicted in Fig. 2, where $\{\xi, \eta\}$ represent the lateral coordinates of the object's position; $(x, y)$ represent the hologram plane. $(x', y')$ represent the plane of the reconstructed image. Therefore, the numerical aperture (NA) of the hologram $NA_{holo}$ and sensor $NA_{sensor}$ are shown in Eq. (5) and Eq. (6), respectively. The smaller NA is chosen as the real NA between $NA_{holo}$ and $NA_{sensor}$, ans is renamed $NA_{real}$.

$$NA_{holo} = \frac{D}{2z}, \tag{5}$$

$$NA_{sensor} = \frac{0.61\lambda}{2\Delta_{pp}}, \tag{6}$$

$$NA_{dhs} = \text{MIN}(NA_{holo}, NA_{sensor}), \tag{7}$$

where $\text{MIN}(\cdot)$ represents choosing the minimum one. The lateral resolution $\Delta LR_{dhy}$ of the digital holographic system is shown in Eq. (8), and the axial resolution $\Delta z'$ of the hologram is shown in Eq. (9) [30-31].

$$\Delta LR_{dhs} = \Delta x' = \Delta y' = \frac{\lambda}{NA_{dhs}}, \tag{8}$$

$$\Delta z' = \frac{\lambda}{(NA_{dhs})^2} \tag{9}$$

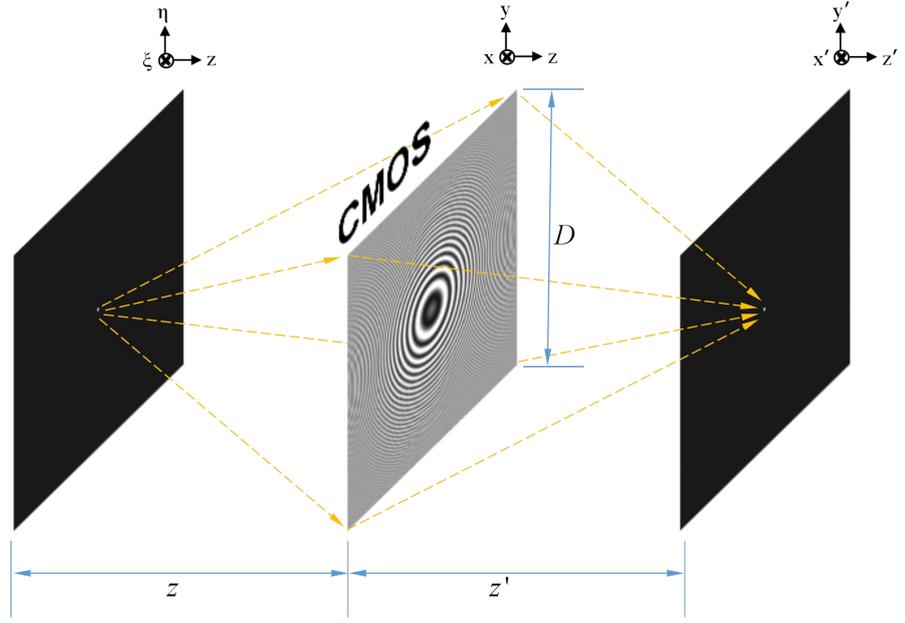

**Figure 2.** Schematic diagram of holographic system for one point.

*3.2. Constrained Intensity for Each Candidate Focused Particle*

Subsequently, we must find out a suitable threshold value to recognize the candidate focused particles in each raw reconstructed image generated from the hologram, in which there are a large number of transparent particles. A flowchart illustrating the procedure of how to find a suitable constrained intensity is shown in Fig. 3. First, a stack of raw reconstructed images $Reim\_all(x,y,n)$ and corresponding gradient images $Reim\_grad\_all(x,y,n)$ is generated. Subsequently, the minimum value along the z-axis is extracted from $Reim\_all(x,y,n)$ to obtain a synthetic minimum intensity image $Reim\_\min(x,y)$. Similarly, the maximum value along the z-axis is extracted from $Reim\_grad\_all(x,y,n)$ to obtain a synthetic maximum gradient image $Reim\_grad\_\max(x,y)$. Then, a threshold in the range of $[v1,v2]$ is sequentially set to binarize $Reim\_\min(x,y)$ to obtain a binarization image $Reim\_\min\_bin(x,y)$. Canny edge detection is applied to obtain the edge of each particle. Afterward, one particle $Pcle(x,y)$ is cropped from $Reim\_\min\_bin(x,y)$, and the corresponding position $Pcle\_grad(x,y)$ in the $Reim\_grad\_\max(x,y)$ is multiplied to obtain the mean value, applying $Grad\_mean = mean(Pcle(x,y).*Pcle\_grad(x,y),"all")$. The threshold is considered to be suitable when $Grad\_mean$ reaches the maximum value. Several other particles are similarly chosen, and their corresponding suitable thresholds similarly calculated. Finally, among these, the minimum threshold is selected; this is the most suitable constrained intensity for the candidate focused particles[32].

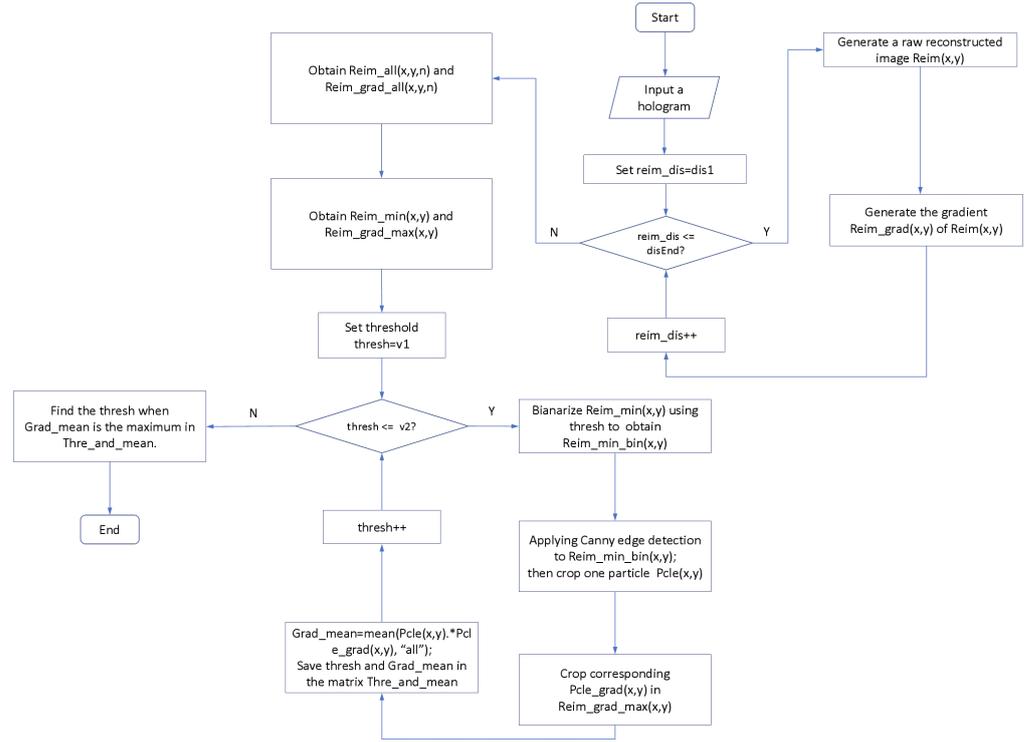

**Figure 3.** Flowchart of procedure to find most suitable constrained intensity for candidate focused particles.

*3.3. Description of Algorithm*

  It is assumed that the hologram is captured using an optical setup, the schematic of which is shown in Fig. 1. The axial resolution is obtained as described in Section 3.1, to set the smallest axial distance by which to recognize a focused particle. A flowchart of the overall procedure is shown in Fig. 4(a), which references process b and process c. Process b, shown in Fig. 4(b), recognizes and saves the information on each candidate focused particle in matrix M, where process c, shown in Fig. 4(c), recognizes the focused particles from matrix M and saves their information in matrix M_new.

  As depicted in Fig. 4(a), the reconstruction distance range $[dis1, dis\_End]$ with a fixed reconstruction depth spacing is settled first, and a series of raw reconstructed images are generated using the back Fresnel propagation method. Subsequently, process b is implemented to obtain the matrix M, in which all candidate focused particles are saved. Finally, process c is implemented to obtain the matrix M_new, in which all of the focused particles are saved.

  In process b, the constrained intensity *fixed_intensity* is calculated as described in Section 3.2 to constrain the intensity of each candidate focused particle. The particle diameters are known and in the range $[min\_dia, max\_dia]$. After Gaussian filtering, Canny edge detection, hole filling, erosion, and dilation of morphological operations are sequentially applied on one raw reconstructed image, a binary image $reim\_bin(x,y)$ with connected regions is obtained. The function *bwlabel* is utilized to extract all of the connected regions (labels), and the number of labels is obtained. Subsequently, all of the labels are traversed. First, the current label *label_num* is extracted in isolation, and its mean intensity (①) and equivalent diameter (②) are calculated using the function *regionprops*. If ① is smaller than *fixed_intensity* and ② is in the range $[min\_dia, max\_dia]$, *label_num* is recognized as a candidate focused particle. Subsequently, its centroid (x, y) (③ and ④), and ⑤ = ① · ② are continually calculated; the reconstruction distance (⑥) and *reim_index* (⑦) are saved; and *focal_status* is set to 0

(⑧). Then, for candidate focused particle *label_num,* the values of ①, ②, ③, ④, ⑤, ⑥, ⑦, and ⑧ are all saved in matrix M. Eventually, we obtain a matrix M containing the information of all candidate focused particles via process b.

In process c, the focused particles from matrix M are extracted, and the information ①, ②, ③, ④, ⑤, ⑥, ⑦, and ⑧ of each focused particle is saved into a new matrix M_new, except that ⑧ *focal_status* is made equal to 1. The axial resolution *axi_resol* is obtained as described in Section 3.1, to set the smallest axial distance by which to recognize a focused particle. The corresponding *axi_slice_num* is also calculated. We obtain the number (*x_num*) of all candidate focused particles from matrix M. Then, we traverse all of these candidate focused particles. During the process, we set *tempi*=1:*x_num*, extract the information of particle *tempi*, and find the index *index_xy* in M wherein the particles whose *x* and *y* axes of the centroid and the *tempi* particle's *x* and *y* axes of the centroid are both less than six pixels, in a series of the reconstructed images, from *tempi* to $tempi + axia\_slice\_num$. Meanwhile, *focal_status* is set to $M(8, index\_xy) = 2$, to indicate that these candidate focused particles in matrix M have already been traversed. Eventually, we find the index *index_best_focal* that satisfies the condition that $M(5, index\_best\_focal)$ is the minimum value in $M(5, index\_xy)$, and *focal_status* (⑧) is set to be equal to $M(8, index\_best\_focal) = 1$. Finally, we obtain a new matrix composed of all of the focused particles. As the candidate focused particles in matrix M are traversed, *focal_status* is made equal to 2, to indicate that the corresponding particle has already traversed; $focal\_status = 0$ indicates the corresponding particle has not yet been traversed, whereas $focal\_status = 1$ indicates that the front *axi_slice_num* reconstructed images of the corresponding particle have already been traversed and it is a focused particle, but the rear *axi_slice_num* reconstructed images of the corresponding particle have not yet been traversed. Therefore, if the *focal_status* of particle *tempi* is equal to 0 or 1, the particle should be traversed. After all particles are traversed in matrix M, the particles whose *focal_status* are equal to 1 are the focused particles. For the focus metric, we propose using the product (⑤) of the mean intensity (①) and equivalent diameter (②), which is easier for processing multiple particles in DH, according to the experimental results.

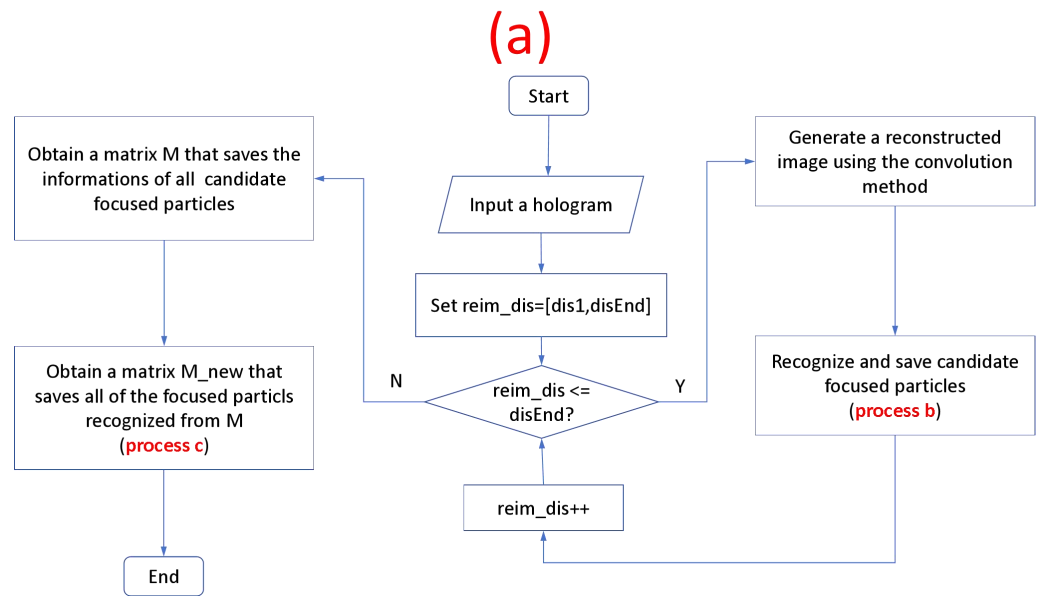

(a)

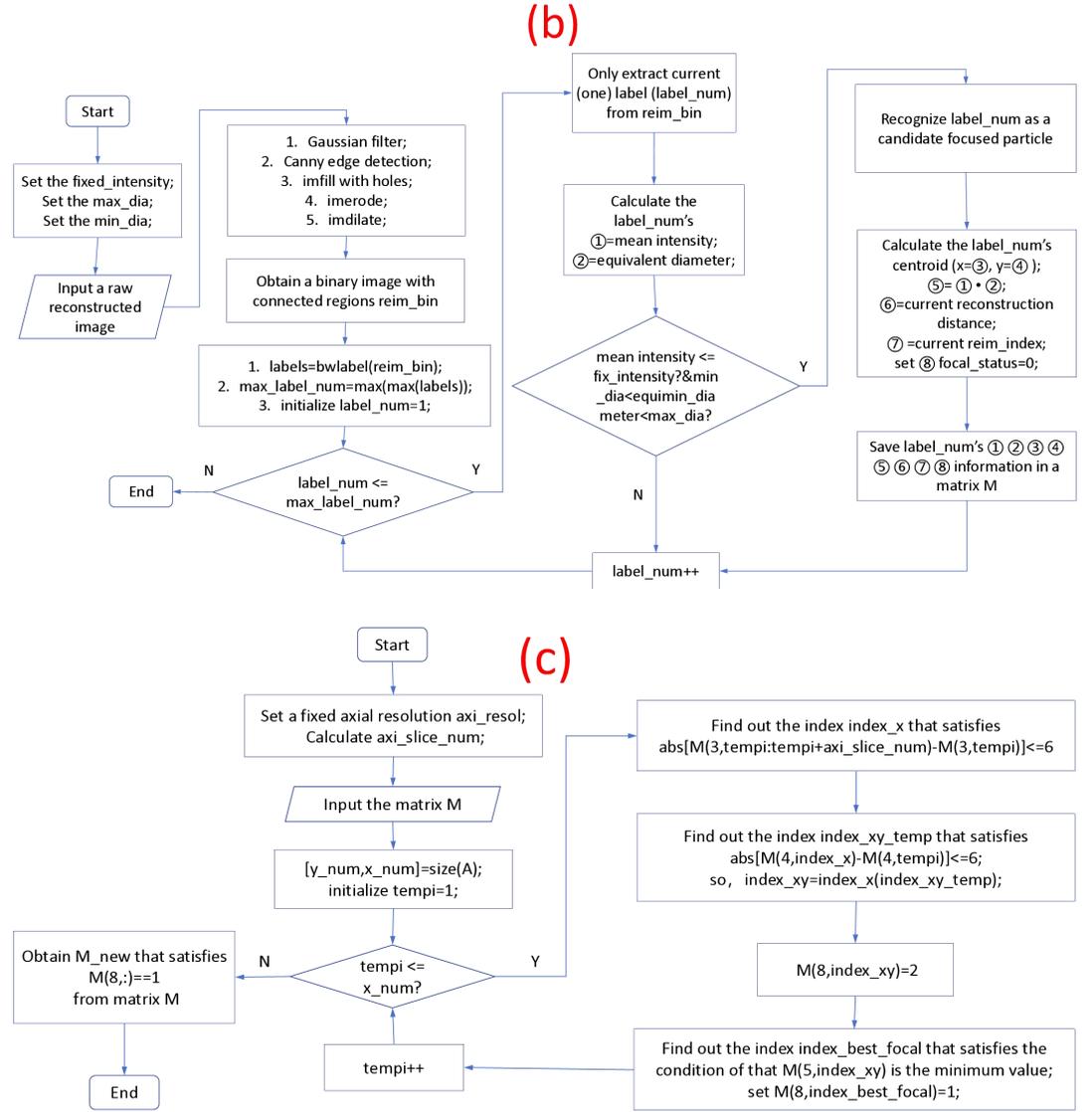

**Figure 4.** Flowcharts of (a) overall process to obtain all focused particles' information, (b) process to obtain and save all candidate focused particles' information in matrix M, and (c) process to determine and save the focused particles in matrix M_new.

## 4. Experimental Results and Analyses

*4.1. Particles in Two Layers*

In this experiment, an in-line digital holographic experimental setup, shown in Fig. 5(a), was used to capture the holograms. The transparent particles were unibead monodispersed polystyrene microspheres with diameters of 50-62 μm and solid content of 2.5% (W/V). First, we used a coherent (green) light source with a wavelength of $\lambda$ = 532 nm to illuminate two layers of particles that are sandwiched between three glass slides, the thickness of each glass slide was about 1 mm. One captured hologram is shown in Fig. 5(b), and the synthetic minimum intensity image, which was the minimum value along the z-axis of all the raw reconstructed images generated in the distance range of [31, 34] mm with a depth spacing equal to 50 μm from this hologram, is shown in the Fig. 5(c). There were a total of 17 particles in the two layers; the particles enclosed by the red circles were placed in one layer, whereas the particles enclosed by green circles were placed in the other layer. The particle centroids, *reim_index*, and reconstruction distances are listed in Table 1. We all obtained 17 particles when the depth spacing was equal to 50 μm (spent 29.0 s and generated 61 reconstructed images),

100 μm (spent 15.2 s and generated 31 reconstructed images), 150 μm (spent 10.5 s and generated 21 reconstructed images), 200 μm (spent 8.4 s and generated 16 reconstructed images), 250 μm (spent 6.9 s and generated 13 reconstructed images), 300 μm (spent 6.2 s and generated 11 reconstructed images), respectively. The reconstruction distances of the particles corresponding these reconstruction depth spacings are depicted in Fig. 6. We can observe that these particles were relatively well-distributed between the two slices.

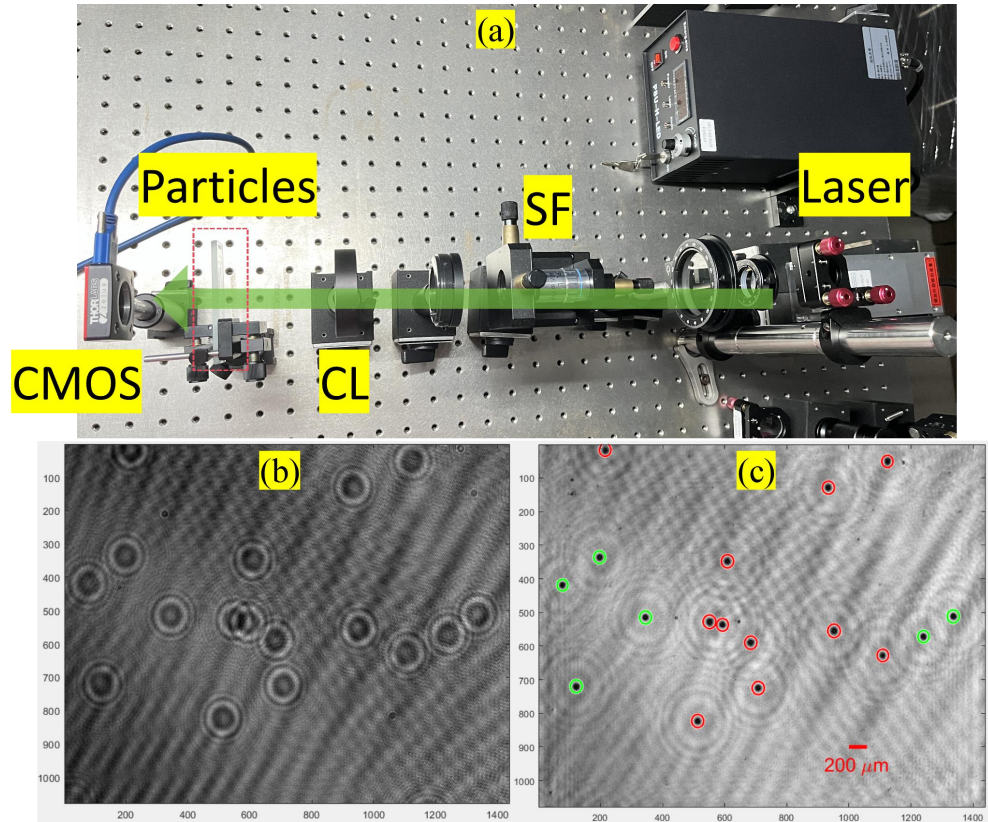

**Figure 5.** (a) Experimental setup (SF: spatial filter, CL: collimating lens), (b) hologram, and (c) synthetic minimum intensity image generated from all raw reconstructed images of hologram in (b).

**Table 1.** Information of Focused Particles in Fig. 5(b) Hologram.

| Particle No. | Centroid [x, y] | Reim_index | Reconstruction distance ( mm) |
|---|---|---|---|
| 1 | [1109, 628] | −43 | 31.85 |
| 2 | [1124, 51] | −42 | 31.90 |
| 3 | [513, 823] | −41 | 31.95 |
| 4 | [684, 590] | −41 | 31.95 |
| 5 | [551, 528] | −40 | 32.00 |
| 6 | [593, 537] | −40 | 32.00 |
| 7 | [609, 348] | −40 | 32.00 |
| 8 | [708, 725] | −40 | 32.00 |
| 9 | [934, 129] | −40 | 32.00 |
| 10 | [952, 555] | −40 | 32.00 |
| 11 | [215, 17] | −35 | 32.25 |
| 12 | [78, 419] | −29 | 32.55 |
| 13 | [197, 336] | −24 | 32.80 |
| 14 | [345, 515] | −22 | 32.90 |
| 15 | [1240, 572] | −22 | 32.90 |
| 16 | [1336, 512] | −22 | 32.90 |
| 17 | [122, 721] | −21 | 32.95 |

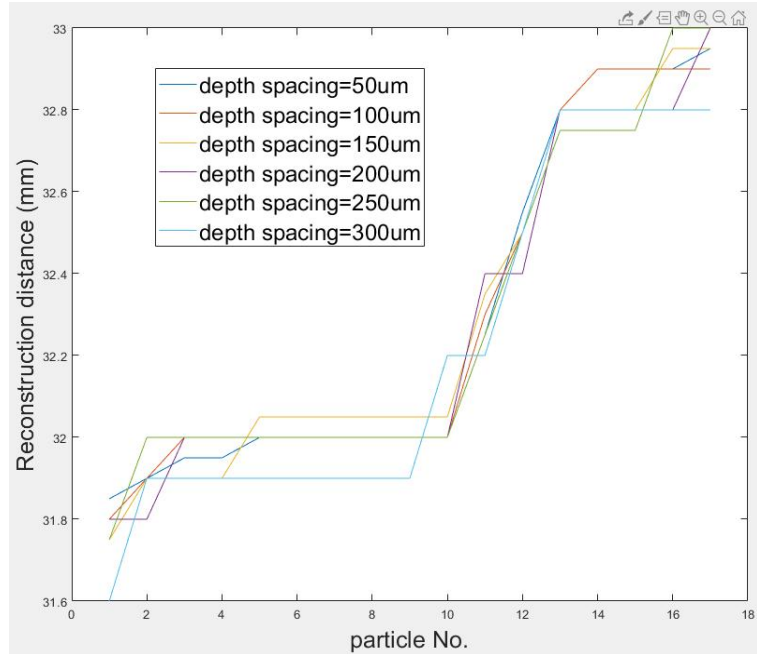

**Figure 6.** Reconstruction distance of each particle corresponding to reconstruction depth spacings equal to 50 μm, 100 μm, 150 μm, 200 μm, 250 μm, 300 μm, respectively.

*4.2. Particles in Cuvette*

We continually captured holograms of large numbers of particles suspended in Milli-Q water that filled 3-mm (the dimension is 12.5*3*45 mm$^3$) and 10-mm (the dimension is 12.5*10*45 mm$^3$) cuvettes, respectively. In this section, only one particle solution, where approximately 3205 particles were seeded per ml, was made. We performed four experiments. In the first one, the 3-mm cuvette filled with the particle solution was placed at 30 mm away from the CMOS camera, and 16 holograms, each of which approximately contained 30 particles, were captured; a hologram sample is shown in Fig. 7(a). We calculated the axial resolution at the distance of 30 mm to be approximately 2.5 mm, and the constrained intensity for each candidate focused particle must be less than 0.39, according to Section 3.2. Using the proposed method, we counted 498 focused particles from the 16 holograms, and obtained a relative error of 3.75%. In the second experiment, we placed the same cuvette with the same particle solution at 40 mm away from the CMOS camera and captured 180 holograms; a hologram sample is shown in Fig. 7(b). We calculated the axial resolution at the distance of 40 mm to be approximately 4.4 mm; however, because the thickness of the cuvette was only 3 mm, we still used the axial resolution of 2.5 mm instead of 4.4 mm, and the same constrained intensity for each candidate focused particle. Using the proposed method, we counted 5796 focused particles from the 180 holograms and obtained a relative error of 7.33%. In the third experiment, we placed the same cuvette with the same particle solution at 60 mm away from the CMOS camera and captured 160 holograms; a hologram sample is shown in Fig. 7(c). We calculated the axial resolution at the distance of 60 mm to be approximately 9.8 mm; however, because the thickness of the cuvette was only 3 mm, we still used the axial resolution of 2.5 mm instead of 9.8 mm, and the same constrained intensity for each candidate focused particle. Using the proposed method, we counted 4841 focused particles from the 160 holograms, and obtained a relative error of 0.85%. In the last experiment, we used a 10-mm cuvette filled with the same particle solution, placed it at 30 mm away from the CMOS camera, and captured 200 holograms; a hologram sample is shown in Fig. 7(d). Since the cuvette was placed 30 mm from the CMOS, we continually set the axial resolution to 2.5 mm and the constrained intensity to be the same as that in the previous experiments. We counted 18851 focused particles from the 200 holograms and obtained a relative error of 5.75%. The results of the four experiments results are presented in detail in Table 2.

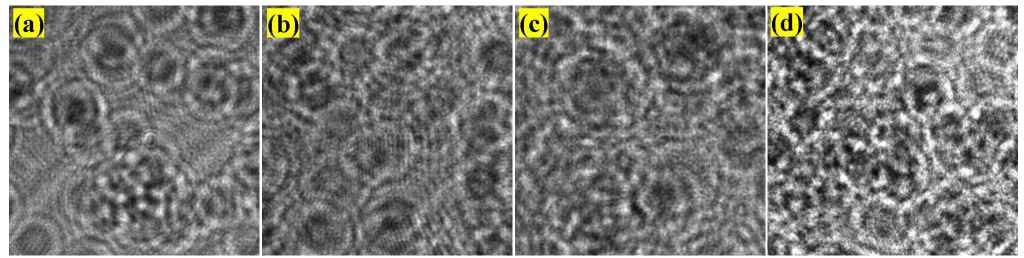

**Figure 7.** Holograms of same particle solutions in (a) 3-mm cuvette placed at 30 mm away, (b) 3-mm cuvette placed at 40 mm away, (c) 3-mm cuvette placed at 60 mm away, and (d) 10-mm cuvette placed at 30 mm away from CMOS, respectively.

**Table 2.** Detailed Results of Four Experiments.

| Experiment No. | 1 | 2 | 3 | 4 |
|---|---|---|---|---|
| Cuvette thickness (mm) | 3 | 3 | 3 | 10 |
| Distance from CMOS (mm) | 30 | 40 | 60 | 30 |
| Axial resolution (mm) | 2.5 | 2.5 | 2.5 | 2.5 |
| Hologram amount | 16 | 180 | 160 | 200 |
| Ground truth of particle amount | 480 | 5400 | 4800 | 20000 |
| Particle amount recognized by proposed method | 498 | 5796 | 4841 | 18851 |
| Deviation | 18 | 396 | 41 | 1149 |
| Relative error (%) | 3.75 | 7.33 | 0.85 | 5.75 |

## 5. Conclusions

We propose an autofocusing method based on morphological analyses and axial resolution to obtain, relatively accurately, the 3D position of each particle, particularly its axial location, and the particle number of a dense transparent particle solution via a hologram. Our proposed method has two components. First, morphological analyses and constrained intensity are used on the raw reconstructed images, from which the information on candidate focused particles is then obtained and saved in one matrix. Second, axial resolution is utilized to obtain the real focused particles from the aforementioned matrix. For the focus metrics, we propose using the product of the mean intensity and equivalent diameter of each candidate focused particle. Eventually, we are able to secure all focused particles for one hologram. In our experiments, we used particles located both at fixed distances and in a particle solution filling a cuvette to examine and verify the proposed method. The deviations of recognized axial locations were in the range of $\pm 0.1$ mm, and relative errors of recognized particle numbers were less than 8%. Therefore, the proposed method is able to rapidly provide relatively accurate ground-truth axial positions for machine learning methods to solve the autofocusing problem that occurs with dense particles and makes it convenient to generate ground-truth datasets for these machine learning methods.

**Funding.** This study was supported by the STU Scientific Research Foundation for Talents and Guangdong University Key Platform [No. 2021GCZX009].

**Disclosures.** The authors declare no conflicts of interest.

**Data availability.** Data underlying the results presented in this paper are not publicly available at this time but may be obtained from the authors upon reasonable request.